\documentclass[pra,aps,amsmath,amssymb,lengthcheck,superscriptaddress,floatfix]{revtex4}
\usepackage{graphicx}
\usepackage{dcolumn}
\usepackage{bm}
\usepackage[usenames,dvipsnames]{color}
\usepackage{psfrag}

\begin{document}

\title{Quantum signatures of self-trapping transition in attractive lattice bosons}

\author{P. Buonsante}
 \affiliation{CNISM, u.d.r. Politecnico di Torino, Corso Duca degli Abruzzi 24, I-10129 Torino, Italy}%
 \affiliation{Dipartimento di Fisica, Politecnico di Torino, Corso Duca degli Abruzzi 24, I-10129 Torino, Italy}%
\author{V. Penna}
 \affiliation{Dipartimento di Fisica, Politecnico di Torino, Corso Duca degli Abruzzi 24, I-10129 Torino, Italy}%
 \affiliation{CNISM, u.d.r. Politecnico di Torino, Corso Duca degli Abruzzi 24, I-10129 Torino, Italy}%
\author{A. Vezzani}
 \affiliation{CNR-INFM, S3, National Research Center, via Campi 213/a, 41100 Modena, Italy}
\affiliation{Dipartimento di Fisica, Universit\`a degli Studi di Parma, V.le G.P. Usberti n.7/A, 43100 Parma, Italy}
\date{\today}

\begin{abstract}
We consider the Bose-Hubbard model describing attractive bosonic particles hopping across the sites of a translation-invariant lattice, and compare the relevant ground-state properties with those of the corresponding symmetry-breaking semiclassical nonlinear theory. The introduction of a suitable measure allows us to highlight many correspondences between the nonlinear theory and the inherently linear quantum theory, characterized by the well-known {\it self-trapping} phenomenon. In particular we demonstrate that the  localization properties and bifurcation pattern  of the semiclassical ground-state  can be clearly recognized at the quantum level. Our analysis highlights a {\it finite-number} effect.  
\end{abstract}
\maketitle

\section{Introduction}
Over the last few years the Bose-Hubbard (BH) model \cite{Fisher_PRB_40_546} has attracted considerable attention, owing to its realization in terms of ultracold atoms trapped in optical lattices \cite{Jaksch_PRL_81_3108,Greiner_Nature_415_39}. 
Among other aspects, the interest in the relation between the BH model with attractive interactions and the nonlinear theory that can be considered its semiclassical counterpart \cite{Bernstein_PhysicaD_68_174,Bernstein_Nonlinearity_3_293,Wright_PhisicaD_69_18} was rekindled \cite{Jack_PRA_71_023610,Buonsante_PRA_72_043620,Oelkers_PRB_75_115119,Javanainen_PRL_101_170405}. 
One of the most striking issues in this respect lies in the symmetry properties of the typical states of the theory.  The translation symmetry of the quantum states appears to be at odds with the spontaneous symmetry breaking of the localized, {\it self-trapped} semiclassical states, although the same regimes can be identified in the two theories \cite{Jack_PRA_71_023610,Buonsante_PRA_72_043620}.
These apparently conflicting features have been reconciled in a recent work \cite{Javanainen_PRL_101_170405} where it is discussed how a {\it single measurement} in a (inherently linear) quantum system can give rise to the localized, symmetry-breaking result typical of the corresponding nonlinear theory. The point in question is illustrated in some detail for  the well known case of a lattice comprising only two sites, and a few further examples are given for a larger lattice, at a single value of the effective parameter governing the system.

Here we extend such analysis in a twofold way. On the one hand, we introduce a suitable notion for the width of a localized state, thereby providing a more quantitative tool for comparing the quantum measurements with the relevant semiclassical predictions. On the other hand, unlike Ref.~\cite{Javanainen_PRL_101_170405}, we do not limit our simulations to the localized regime, but we perform a systematic comparison, exploring a range of effective parameters including the semiclassical bifurcation of the system ground state \cite{Eilbeck_PhysicaD_16_318,Smerzi_PRL_89_170402}.
The analysis of the {\it localization width} shows that the semiclassical delocalization transition corresponds to a crossover in the quantum system, which highlights a {\it finite-population} effect. The sharp transition is recovered only in the limit of infinite boson filling, where the classical results are exact.

Also, the range of effective parameters and lattice sizes we explore includes those highlighting the change in the nature of the bifurcation occurring at the semiclassical level when passing from five-site to six-site lattices \cite{Buonsante_JPB_39_S77,Buonsante_PRE_75_016212}.  As we discuss in the following, a signature of this qualitative change is apparent at the quantum level. 

The layout of this paper is the following. In Section~\ref{S:system} we review the second-quantized model for attractive bosons hopping across the site of a lattice and the nonlinear theory representing its semiclassical counterpart.
We discuss the localization/delocalization transition occurring in the ground-state of the latter, revisiting the results discussed in Ref.~\cite{Javanainen_PRL_101_170405}. Also, we provide a quantitative measure for localization, which is readily extended to the quantum theory.

 Section~\ref{results} contains our results, which are based on the numerical calculation of the ground-state of the quantum system. This is obtained by means of Lanczos diagonalization and {\it population} quantum Monte Carlo simulations. We start with a detailed analysis of the specific case of a three-site lattice (trimer), for which  the localization/delocalization transition can be appreciated by a direct visualization of the structure of the ground-state. We then make a systematic analysis of the localization width for several lattice sizes and boson fillings, highlighting the agreement with the semiclassical result for large fillings. This also shows that the qualitative change in the bifurcation pattern of  the classical theory taking place for lattices containing more than five sites is recovered at the quantum level.

In Section \ref{S:CS} we discuss the connection between the quantum and classical theory based on the representation of the quantum ground-state as a su$(L)$ coherent state \cite{Buonsante_PRA_72_043620} (also known as {\it Hartree wave function} \cite{Wright_PhisicaD_69_18}).
This allows us to employ the semiclassical ground state to construct an approximation to the quantum ground which includes some of the {\it finite population} effects mentioned above. 

\section{The system}
\label{S:system}
We consider a simple Bose-Hubbard model 
\begin{equation}
\label{BH}
\hat H = -\sum_{j=0}^{L-1} \left[\frac{U}{2} \left(\hat a_j^\dag\right)^2 \hat a_j^2 + J \left(\hat a_j^\dag \hat a_{j+1}+ \hat a_{j+1}^\dag \hat a_{j}\right)\right],
\end{equation}
with attractive interactions, $U,\,J>0$.
The operators  $\hat a_j^\dag$ and $\hat a_j$  create and destroy  bosons at lattice site $j$, respectively. The lattice comprises $L$ sites and is periodic, so that 
 site label $j=L$ is to be identified with $j=0$. The (positive) parameters $U$ and $J$  account for the relative strength of the {\it interaction} and {\it kinetic} term, respectively.

The BH Hamiltonian clearly commutes with the total number of bosons in the system $\hat N=\sum_j \hat n_j$, with  $\hat n_j = \hat a_j^\dag \hat a_j$. This means that the eigenstates of Eq.~\eqref{BH} are characterized by a well defined total number of bosons $N$, i.e. that $H$ can be studied within a fixed-number subspace of the infinite Hilbert space. The size of this fixed-number subspace is finite, $s = \binom{N+L-1}{L}$,  but it becomes computationally prohibitive already for modest lattice sizes and boson populations.

Hamiltonian \eqref{BH} is clearly translation invariant. By virtue of the Bloch theorem, its eigenstates --- and in particular its ground state $|\Psi\rangle$ --- are delocalized over the entire lattice. This means $n_j=\langle\Psi|\hat n_j|\Psi\rangle = \langle\Psi|\hat n_\ell |\Psi\rangle =n_\ell$ for any $j$ and $\ell$.

The  semiclassical nonlinear theory corresponding to the BH model can be obtained by changing the site operators  into C-numbers, $ \hat a_j^\dag \leadsto \psi_j$, whose square modulus has a natural interpretation as the local boson population, $\hat n_j \leadsto  |\psi_j|^2$. The semiclassical Hamiltonian thus modeled onto Eq. \eqref{BH} results in a dynamics governed by the so-called {\it discrete nonlinear Schr\"odinger equations} or {\it discrete self-trapping } (DST) {\it equations},
\begin{equation}
\label{DST}
i \dot \psi_j = - U |\psi_j|^2 \psi_j - J \left(\psi_{j+1} + \psi_{j-1}\right)
\end{equation}
where we set $\hbar = 1$ \cite{Eilbeck_PhysicaD_16_318}. The stationary solutions $\psi_j(t)= e^{-i \omega t} \phi_j$ obey the fixed point equations
\begin{equation}
\label{DSTfp}
\omega \phi_j = - U |\phi_j|^2 \phi_j - J \left(\phi_{j+1} + \phi_{j-1}\right)
\end{equation}

Similar to the quantum case, the semiclassical Hamiltonian, as well as equations of motion \eqref{DST} and \eqref{DSTfp}, are translation invariant. However, their nonlinear nature  allows of symmetry-breaking solutions. 
In particular, it is easy to check that Eq.~\eqref{DSTfp} always has a uniform solution, $\phi_j = \sqrt{N/L}$, independent of the value of the parameters $U$ and $J$. Its frequency and energy are $\omega = -U N/L-J$ and $E = -U N/2L$, respectively,  the latter being the lowest possible energy only for  sufficiently large values of the effective parameter 
\begin{equation}
\label{tau}
\tau = \frac{J}{U N},\qquad  \tau > \tau_{\rm d}
\end{equation}
If, conversely, the effective interaction energy $U N$ prevails over the hopping amplitude $J$, so that $\tau < \tau_{\rm d}$, the lowest-energy solution of Eq. \eqref{DSTfp}  is localized about one lattice site, thus breaking the translation invariance.  This phenomenon is referred to as ({\it discrete}) {\it self-trapping} \cite{Eilbeck_PhysicaD_16_318}. In general, a second critical value
\begin{equation}
\label{taus}
\tau_{\rm s}= \frac{1}{2 L \sin^2 \frac{\pi}{L}} \leq \tau_{\rm d}
\end{equation}
can be recognized for the uniform solution of Eq.~\eqref{DSTfp} such that
for $\tau> \tau_{\rm s}$ the uniform solution is dynamically stable, whereas it is unstable for $\tau< \tau_{\rm s}$ \cite{Smerzi_PRL_89_170402}. 

For sufficiently large lattice sizes, $L\geq 6$, the equality applies in Eq.~\eqref{taus}. That is, the uniform solution becomes simultaneously stable {\it and} the ground state of the system as soon as $\tau> \tau_{\rm s} = \tau_{\rm d}$. On smaller lattices the strict inequality applies, $\tau_{\rm s} < \tau_{\rm d}$, and the low-energy solutions of \eqref{DSTfp} exhibit a more complex bifurcation pattern \cite{Buonsante_JPB_39_S77,Buonsante_PRE_75_016212,Note1}. 

A signature of the localization occurring at the semiclassical level for large values of the interaction can be recognized also at the quantum level.
The quantum ground-state can be seen a symmetric superposition of $L$ states, each localized at one of the $L$ sites of the lattice and closely resembling the symmetry-breaking semiclassical ground-state. At the boson fillings $N/L$ corresponding to Hamiltonians that can be analyzed by means of Lanczos diagonalization, this resemblance is strong only for very small values of the effective parameter $\tau$, while it is washed out by  quantum fluctuations  at larger $\tau$'s  \cite{Buonsante_PRA_72_043620}.

In Ref.~\cite{Javanainen_PRL_101_170405} the connection between the uniform quantum ground state and its localized semiclassical counterpart is further discussed. There it is remarked that  the experimental measurement of the observable $\hat {\mathbf{n}}= (\hat n_1,\hat n_2,\cdots,\hat n_L) $ is likely to produce an outcome in strong agreement with the semiclassical result. Indeed such a measurement selects a Fock state of the direct space, i.e. an eigenstate of $\hat {\mathbf n}$,
\begin{equation}
\label{nu}
|\vec{\nu}\rangle=|\nu_1,\nu_2,\cdots,\nu_L\rangle = \prod_{j=1}^L \frac{(a_j^\dag)^{\nu_j}}{\sqrt{\nu_j!}} |0\rangle\!\rangle,
\end{equation}
where $|0\rangle\!\rangle$ is the vacuum of the theory, i.e. $a_j |0\rangle\!\rangle$ for all $j=1,2,\ldots,L$.
Each Fock state is selected with  a probability distribution given by $P(\vec{\nu}) = |c_{\vec{\nu}}|^2$, where the  $c_{\vec{\nu}}$'s are the coefficients in the expansion 
\begin{equation}
\label{qGS}
|\Psi\rangle  = {\sum_{\vec{\nu}}}' c_{\vec{\nu}} |\vec{\nu}\rangle
\end{equation}
The prime on the summation symbol signals that the Fock states involved in the expansion belong to the same fixed-number subspace, $\sum_j \nu_j = N$.

The Authors of Ref.~\cite{Javanainen_PRL_101_170405} first of all recall that for a two-site lattice, $L=2$, the  probability distribution $P(\vec{\nu})$ is double peaked when $\tau<\tau_{\rm d}$ \cite{Ho_JLTP_135_257,Zin_EPL_83_64007}	, and observe that the mirror-symmetric Fock states corresponding to the peak probabilities reproduce the populations of the symmetry-breaking semiclassical ground states: $(\nu_1,\nu_2) = (|\psi_1|^2,|\psi_2|^2)$,  $(\nu_1,\nu_2) = (|\psi_2|^2,|\psi_1|^2)$. Also, by exactly diagonalizing Hamiltonian \eqref{BH} they show that the width of the probability peaks decrease with increasing total number of bosons. 

The Authors' observation is further illustrated by considering the case of $N=256$ bosons on a lattice comprising $L=16$ sites. A few Fock states $|\vec{\nu}\rangle$ are {\it sampled} from a quantum Monte Carlo calculation and, after suitable translations, graphically compared with the semiclassical result for the local populations $(|\psi_1|^2, |\psi_2|^2,\cdots,|\psi_L|^2)$. The importance-sampled quantum configurations agree reasonably well with the semiclassical result, although  quantum fluctuations are clearly recognizable. 

In summary, Ref.~\cite{Javanainen_PRL_101_170405} demonstrates how a measurement in an inherently linear quantum theory could produce a typical outcome of a nonlinear theory, i.e. a localized {\it soliton-like} ground-state.

In the following we further investigate the connection between the ground-state of Eqs. \eqref{BH} and \eqref{DSTfp}, making the comparison more quantitative with the aid of a suitable observable. Also we explore a range of effective parameters $\tau$ including the  localization/delocalization threshold $\tau_{\rm d}$, and discuss the signature of this semiclassical transition at the quantum level.  In particular we illustrate that  the difference in the semiclassical bifurcation pattern distinguishing small lattices from those comprising more than six sites \cite{Buonsante_JPB_39_S77,Buonsante_PRE_75_016212} is apparent also in the quantum data.
\subsection{Soliton width}
\begin{figure}
\begin{centering}
\includegraphics[width=8.5 cm]{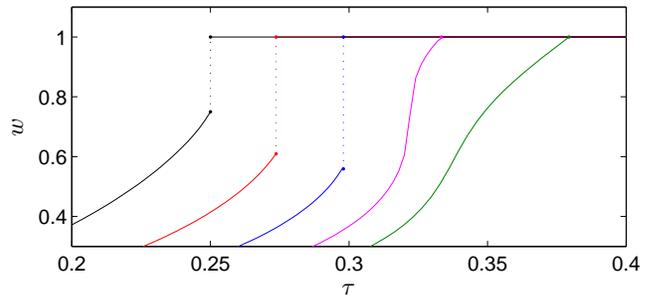}
\caption{\label{scW} Width of the semiclassical ground-state according to Eq.~\eqref{w2} for some lattice sizes, from left to right $3 \leq L \leq 7$. Notice that the transition from a localized ($w<1$) to a uniform state ($w=1$) is discontinuous for $L<6$ \cite{Buonsante_JPB_39_S77,Buonsante_PRE_75_016212}. The dotted lines at the critical values $\tau_{\rm d}$ are guides to the eye. }
\end{centering}
\end{figure}

As we recall above, for $\tau < \tau_{\rm d}$ the ground-state of the DST equations \eqref{DSTfp} becomes localized at one of the lattice sites, assuming a {\it soliton-like} density profile. Since the soliton peak can be localized at any of the $L$ lattice sites, the ground-state is $L$-fold degenerate. This spontaneous symmetry breaking entails from the nonlinear nature of Eq.~\eqref{DSTfp}. After recalling that we are considering a cyclic lattice, i.e. periodic boundary conditions, we can estimate the (square) width of a localized solution of Eq.~\eqref{DSTfp} as
\begin{equation}
\label{w2}
w(\vec{n}) = \sum_{j=1}^L \frac{n_j}{N}\left[\left(x_j^2 - x_{\rm cm}^2\right)+\left(y_j^2 - y_{\rm cm}^2\right)\right],
\end{equation}
where $x_j = \cos \frac{2\pi}{L} j$ and $y_j = \sin \frac{2\pi}{L} j$ are the coordinates of the $j$-th site of the ring lattice \cite{note3}, $n_j = |\psi_j|^2$ is the boson population at that site and
\begin{equation}
x_{\rm cm}(\vec{n}) = \sum_{j=1}^L \frac{n_j}{N} x_j, \quad y_{\rm cm}(\vec{n}) = \sum_{j=1}^L \frac{n_j}{N} y_j
\end{equation}
are the coordinates of the center of mass of the boson distribution.
When the semiclassical solution is uniform, $n_j = |\psi_j|^2 = N/L$, the center of mass is at the center of the ring lattice, $x_{\rm cm} = y_{\rm cm}=0 $ and the width attains the maximum possible value, $w=1$.
At the opposite limit, the center of mass coincides with the lattice site at which the entire boson population is confined. In this situation the width attains its minimum value, $w=0$. The  width of the lowest solution to Eq.~\eqref{DSTfp} is plotted in Fig.~\ref{scW} for some small lattices, $3\leq L \leq 7$. Note that the different bifurcation pattern characterizing the smaller lattices \cite{Buonsante_JPB_39_S77,Buonsante_PRE_75_016212}, $L<6$, is mirrored in the discontinuos character of the width of the ground state. In general an increase of $\tau<\tau_{\rm d}$ results in an increase of the width of the localized state. However, the  delocalization is attained continuously at $\tau=\tau_{\rm d}$ only for $L\geq 6$. For smaller lattice sizes the width of the localized ground state has an upper bound smaller than 1, and delocalization is attained {\it catastrophically} as $\tau>\tau_{\rm d}$. 

As we discuss above, a measurement of the quantum observable $\hat{\mathbf{n}}$ selects a Fock state, i.e. a set of integer occupation numbers $\nu_j = \langle\hat n_j\rangle $, with probability $P(\vec{\nu}) = |c_{\vec{\nu}}|$ defined by Eq.~\eqref{qGS}. Eq.~\eqref{w2} lends itself to the estimate of the width of the selected Fock state as well, provided that $\nu_j$ is plugged in $n_j$, instead of $|\psi_j|^2$.
Taking into account that $x_j^2+y_j^2=1$ and that the occupation numbers in the Fock states add up to  $N$, after a few manipulations Eq.~\eqref{w2} becomes
\begin{eqnarray}
w(\vec{\nu}) &=& 1-\frac{1}{N^2}\sum_{j\ell} \cos\left[\frac{2\pi}{L}(j-\ell)\right] \nu_j \nu_\ell \nonumber\\
&=& 1-\frac{1}{N^2}\sum_{j\ell} \cos\left[\frac{2\pi}{L}(j-\ell)\right] \langle \vec{\nu}|\hat n_j \hat n_\ell|\vec{\nu}\rangle  
\end{eqnarray}
Therefore, the average of a large number of measurements of $w$ tends to the quantum observable 
\begin{eqnarray}
\overline{w} &=& {\sum_{\vec{\nu}}}' P(\vec{\nu}) w(\vec{\nu})\nonumber\\
&=&1-
\frac{1}{N^2}\sum_{j\ell} \cos\left[\frac{2\pi}{L}(j-\ell)\right]\left\langle \hat n_j \hat n_\ell 
\right\rangle \nonumber\\
\label{aqW}
&=& 1-\left\langle   \frac{1}{2} \left[\hat S\left(\frac{2\pi}{L}\right)+\hat S\left(-\frac{2\pi}{L}\right)  \right] \right\rangle
\end{eqnarray}
where 
\begin{equation}
\label{ssf}
\hat  S(q) = \frac{1}{N^2} \sum_{j\ell} e^{i q(j-\ell)} \hat n_j \hat n_\ell .
\end{equation}
We note that the quantity in Eq.~\eqref{ssf}  is formally similar to the {\it static structure factor} for a 1D linear lattice \cite{Roth_PRA_68_023604,Note2}.	
\section{Results}
\label{results}
In the following we show that the average width of a quantum ground state, as estimated by Eq.~\eqref{aqW}, reproduces the semiclassical results shown in Fig.~\ref{scW}, the agreement improving as the boson population in the lattice is increased. We thereby make the analysis of Ref.~\cite{Javanainen_PRL_101_170405} more systematic and quantitative. Furthermore, we do not limit our investigation to the region where the semiclassical solution is localized, but explore the transition region as well.  We do this for relatively small lattices,  but  consider one example, the six-site lattice, for which the bifurcation pattern at the transition is qualitatively similar to larger lattices \cite{Buonsante_JPB_39_S77,Buonsante_PRE_75_016212}. 

In order to evaluate the quantity in Eq.~\eqref{aqW} we need to obtain the ground-state of Hamiltonian \eqref{BH}. We achieve this by means of two different numerical approaches. When the size of the relevant Hilbert space is sufficiently small, we employ a {\it Lanczos} diagonalization algorithm. Otherwise, we resort to a stochastic method, the so-called {\it population} quantum Monte Carlo algorithm \cite{Iba_TJSAI_16_279}. In both cases we exploit the symmetry granted by the commutator  $[\hat H,\hat N]$ by working in the canonical ensemble, i.e. by considering only the occupation-number Fock states $|\vec{\nu}\rangle$ relevant to a given total number of bosons, $N=\sum_j \nu_j$. In order to  reduce further the size of the matrix to be analyzed by the Lanczos algorithm, we also take advantage of the translation symmetry of the system. We first of all gather the fixed-number Fock states into equivalence classes determined by lattice translations, which allow to define the reduced basis:
\begin{equation}
\label{tiF}
|\vec{\nu}_*\rangle = \frac{1}{\sqrt{{\cal N}_{\vec{\nu}}}} \sum_{j=1}^{{\cal N}_{\vec{\nu}}} {\hat D}^{\frac{L}{{\cal N}_{\vec{\nu}}}\,j} |\vec{\nu}\rangle
\end{equation}
In the r.h.s. of this equation $|\vec{\nu}\rangle$ represents any member of an equivalence class, which determines all of the other ${\cal N}_{\vec{\nu}}-1$ members of the same class through the {\it displacement operator} $\hat D$ such that $\hat D \hat a_j \hat D^{-1} = \hat a_{j+1}$. The number ${\cal N}_{\vec{\nu}}$ of states in a class is in general a    divisor of the lattice size $L$. 
The size of the reduced basis comprising states of the form \eqref{tiF} is the same as the size of the subspace formed by the  {\it quasimomentum} Fock states $|\vec{q}\rangle$ such that $\sum_{k=1}^L k \,q_k= \kappa\, L$, with $\kappa \in {\mathbb{Z}}$ i.e. roughly a factor $L$ smaller than the fixed-number Fock space.

Before discussing the results for the average quantum width, Eq.~\eqref{aqW}, we analyze the ground state of three-site lattice in some detail.
\subsection{The trimer}
\begin{figure}
\begin{centering}
\includegraphics[width=8.5 cm]{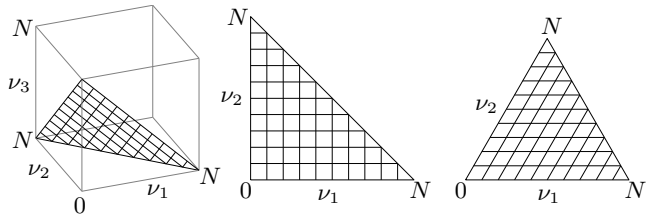} 
\caption{\label{trif1} Representation of the probability distribution $P(\vec{\nu})$. Left: the dark lines demonstrate the (flat) surface of the occupation-number Fock space relevant to a trimer containing $N$ bosons. Center: the same surface as viewn from direction $(0 0 1)$, i.e. projected onto the $(\nu_1,\nu_2)$ plane. Right: the same surface as in the rightmost panel as viewn from the direction normal to it, $(1 1 1)$. Notice that this view highlights the threefold symmetry of the domain of the distribution probability.	}
\end{centering}
\end{figure}
For lattices comprising just three sites the probability distribution $P(\vec{\nu})$ for the Fock state $|\vec{\nu}\rangle=|\nu_1,\nu_2,\nu_3\rangle$ selected in an experiment measuring $\hat {\mathbf n}=(\hat n_1,\hat n_2, \hat n_3)$ can be conveniently represented as  a two dimensional density plot. This is made possible by the total number conservation,  which constrains one of the occupation numbers, say $\nu_3$, to a value depending linearly on the two remaining occupation values $\nu_3=N-\nu_1-\nu_2$. One can therefore regard the probability distribution as a function of the latter alone, $P(\nu_1,\nu_2,N-\nu_1-\nu_2)$.

The portion of the occupation-number Fock space relevant to a trimer containing $N$ boson is illustrated in the leftmost panel of Figure \ref{trif1}. The same surface as seen from two different points of view is shown in the remaining two panels. In the central panel the point of view is on the direction $(0,0,1)$, which results in a projection onto the $(\nu_1,\nu_2)$ plane. In the leftmost panel the point of view is on the direction $(1,1,1)$ normal to the surface under investigation. This highlights the three-fold symmetry of the surface, which is not apparent in the previous view. Since we are interested in the symmetry of the system, we  adopt the second point of view when representing the probability distribution $P(\vec{\nu})$.

\begin{figure}
\begin{centering}
\begin{tabular}{cc}
\includegraphics[width=4.0 cm]{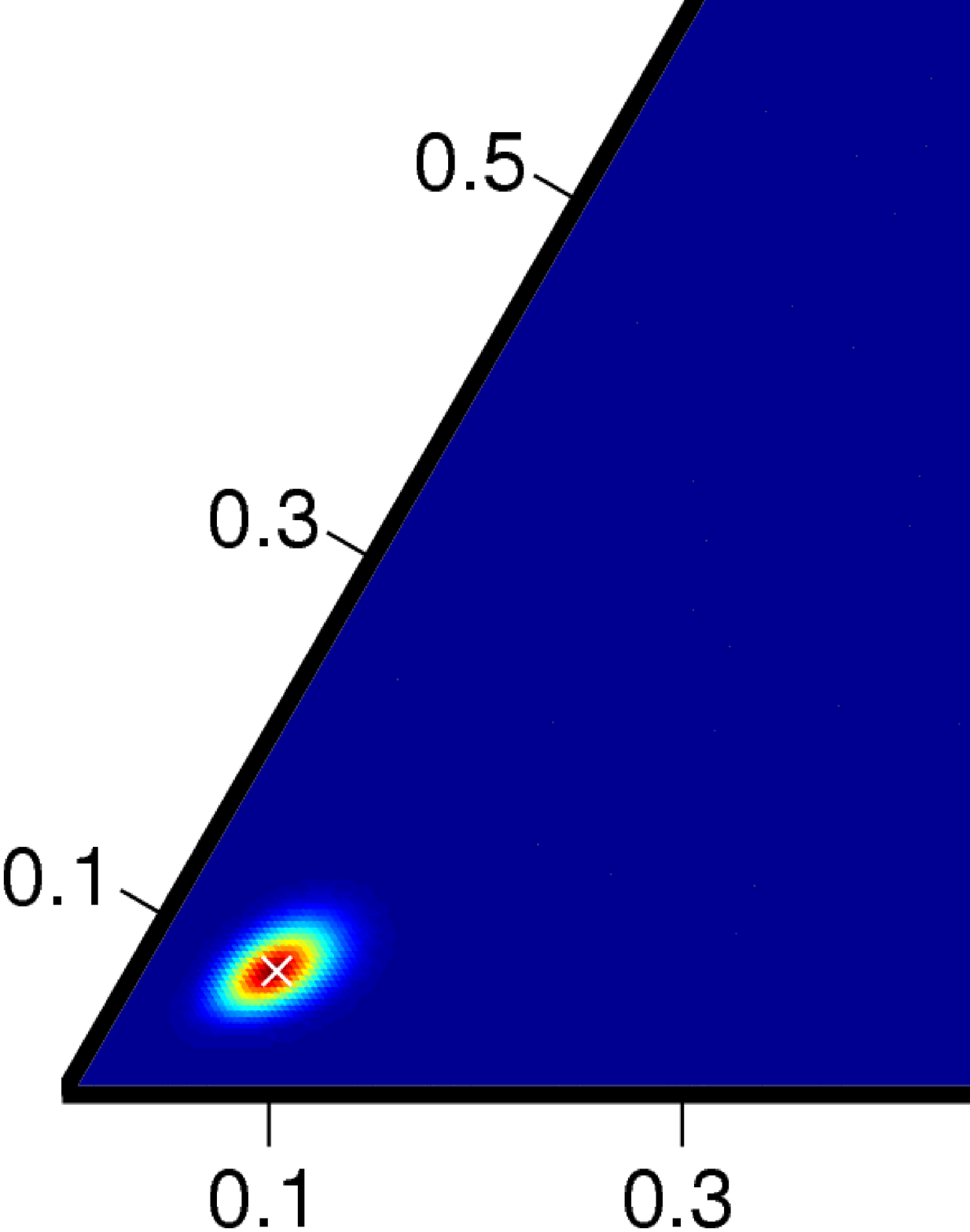} & 
\includegraphics[width=4.0 cm]{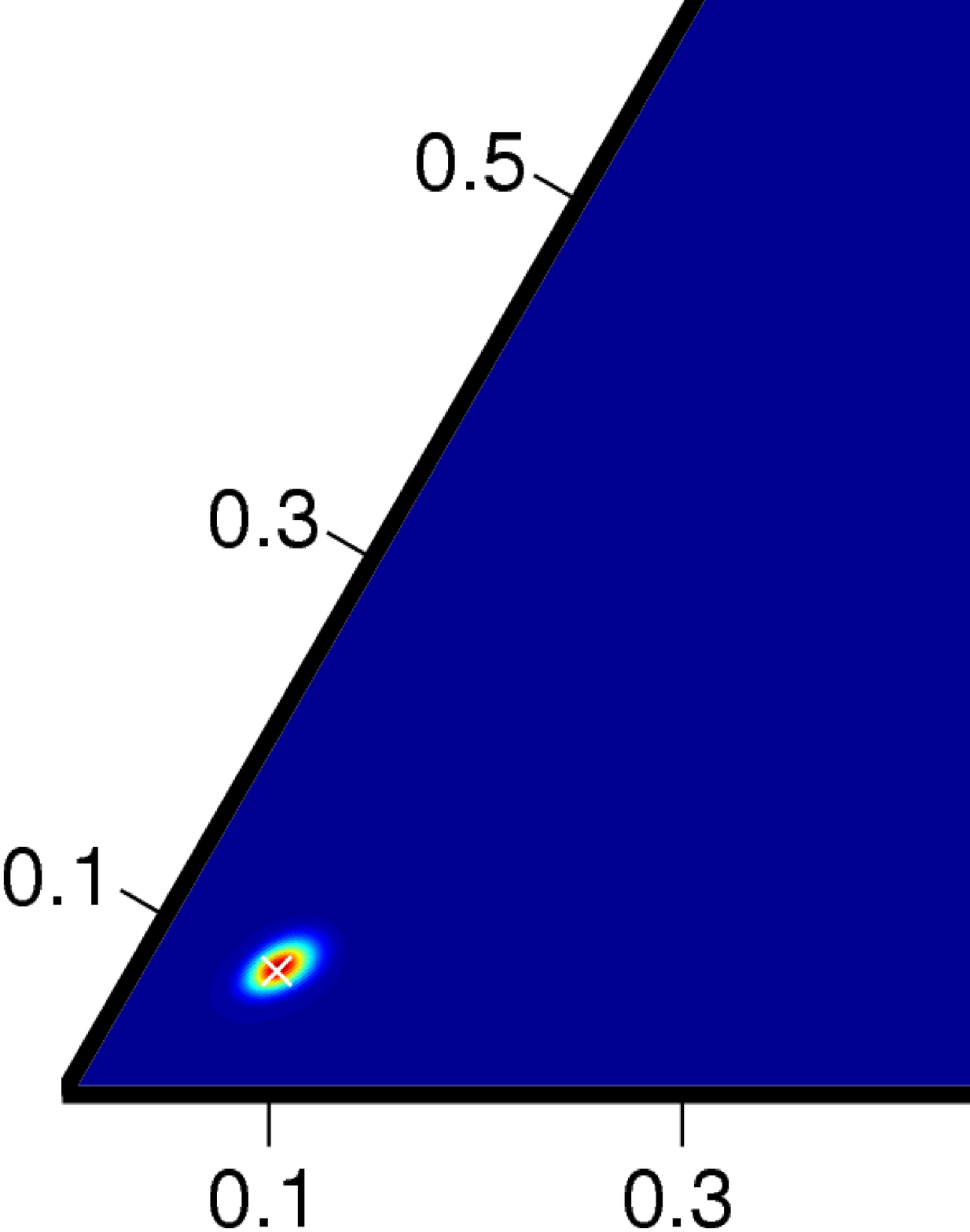}  \\
\includegraphics[width=4.0 cm]{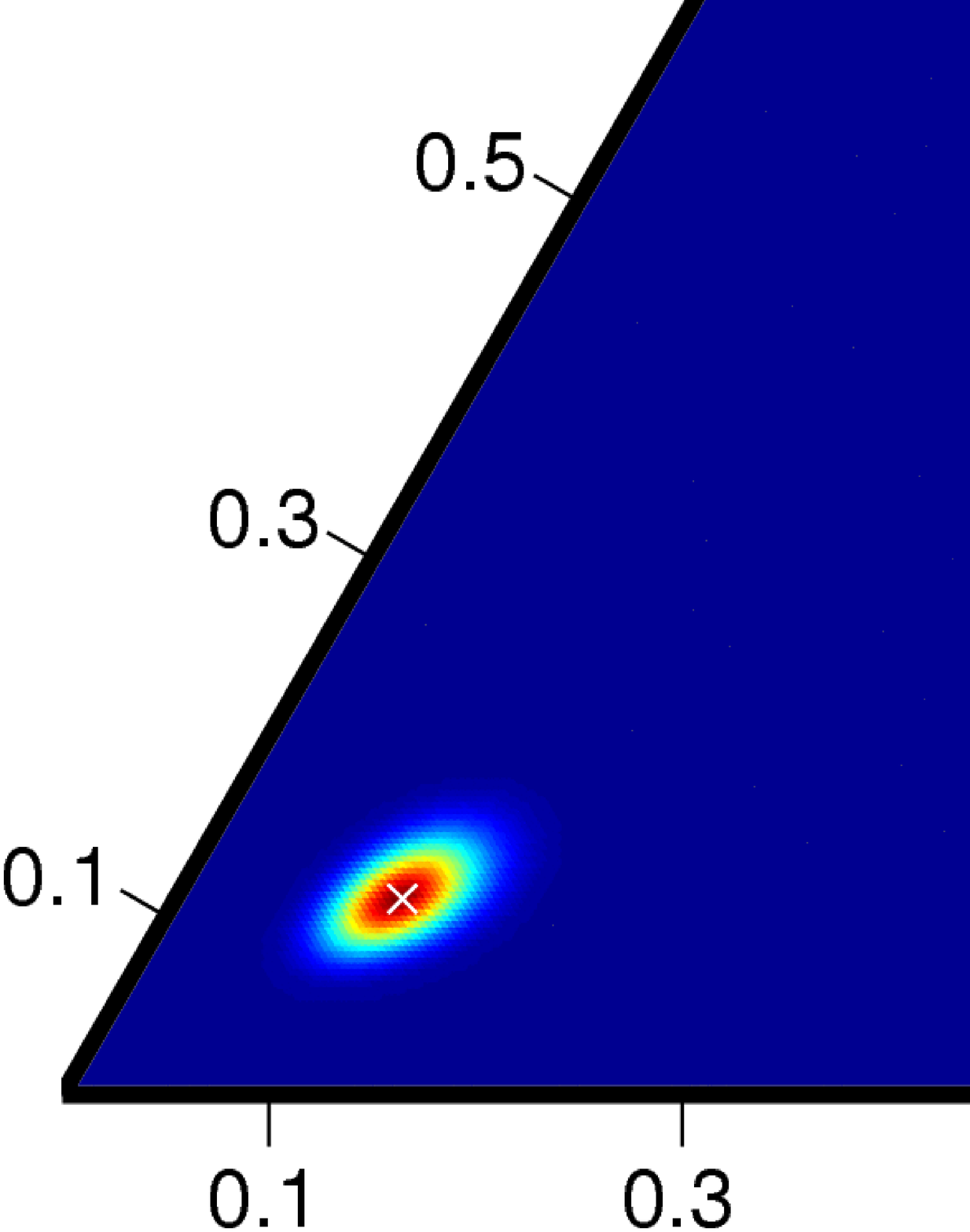} & 
\includegraphics[width=4.0 cm]{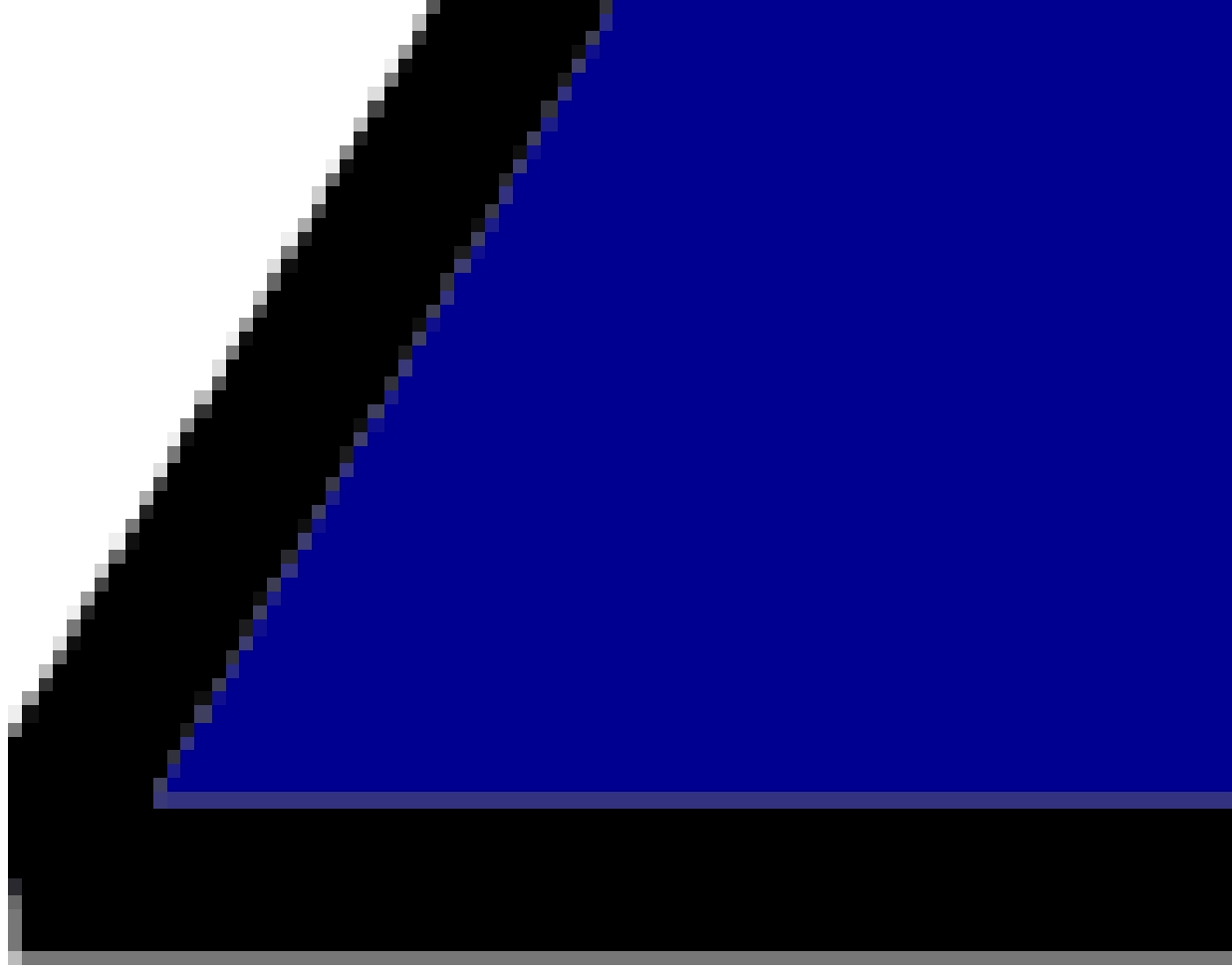} \\ 
\includegraphics[width=4.0 cm]{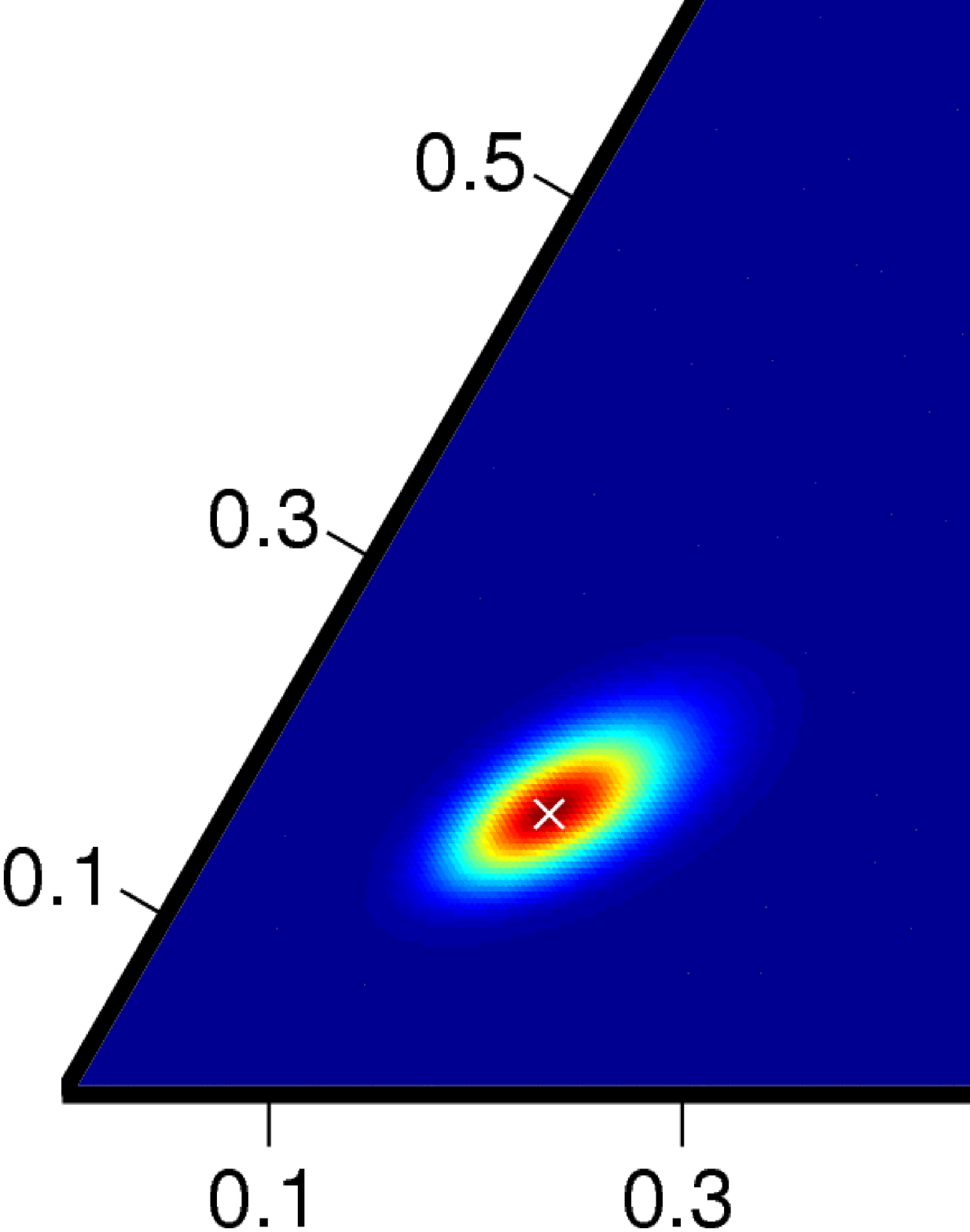} & 
\includegraphics[width=4.0 cm]{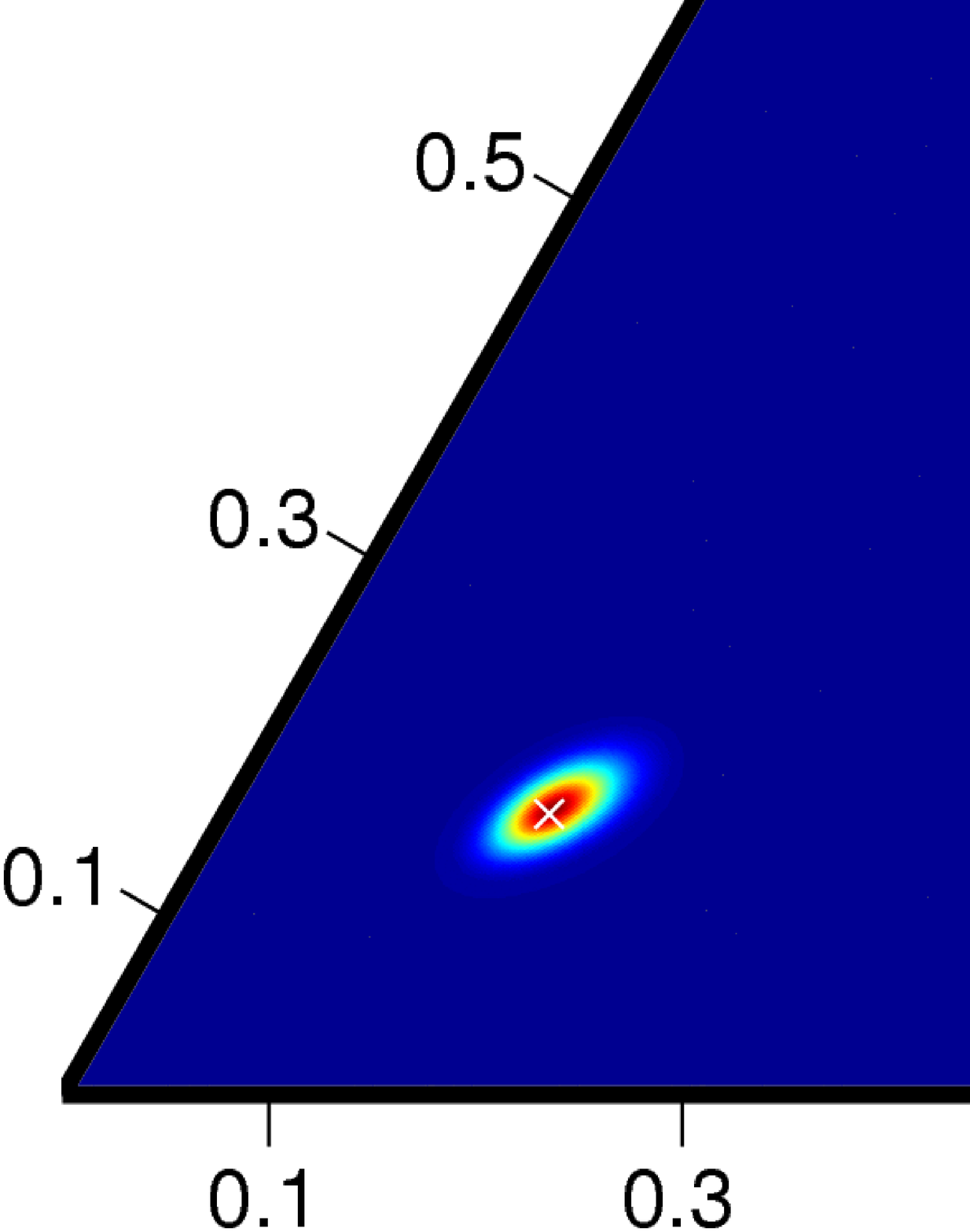} \\
\includegraphics[width=4.0 cm]{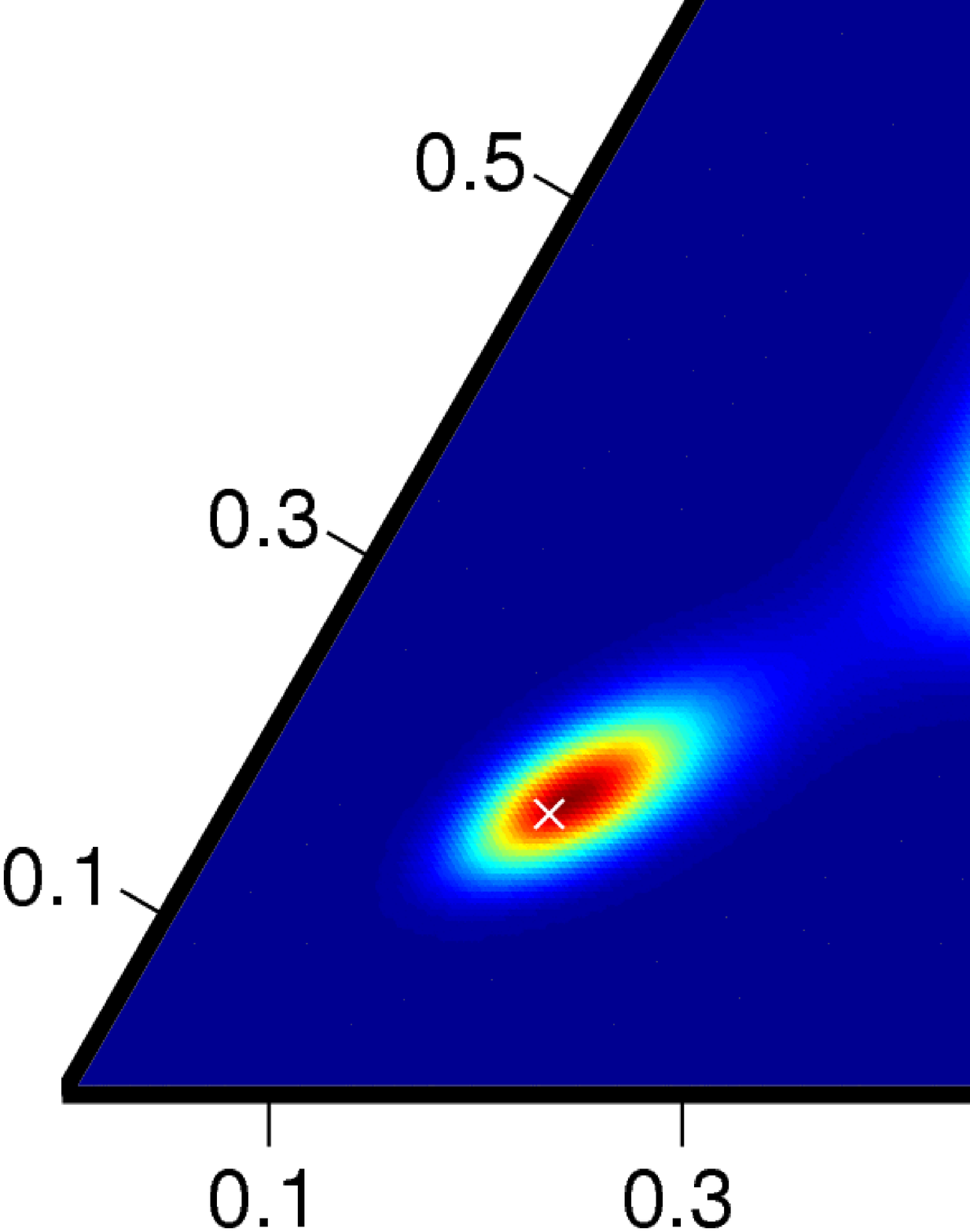} & 
\includegraphics[width=4.0 cm]{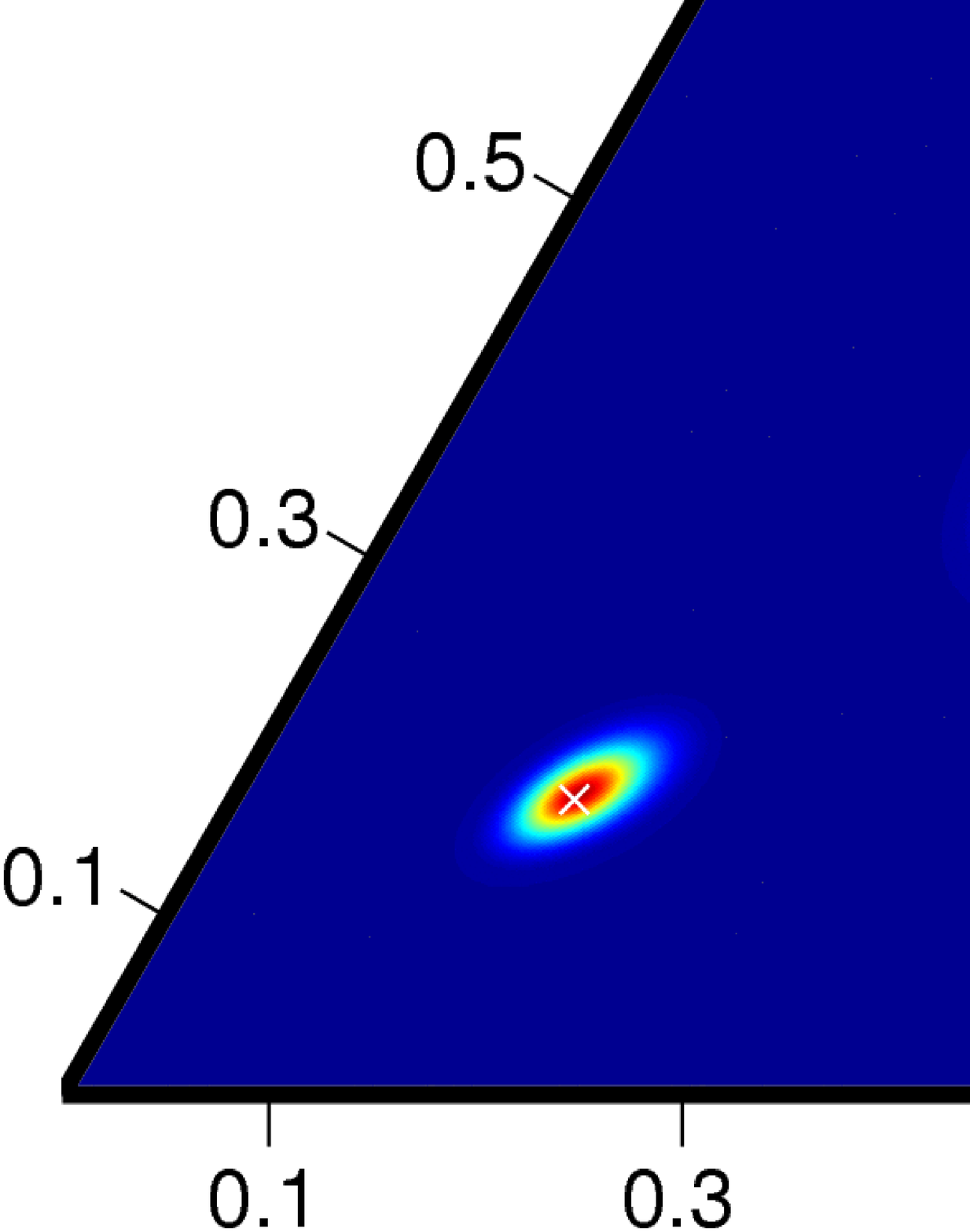} \\
\includegraphics[width=4.0 cm]{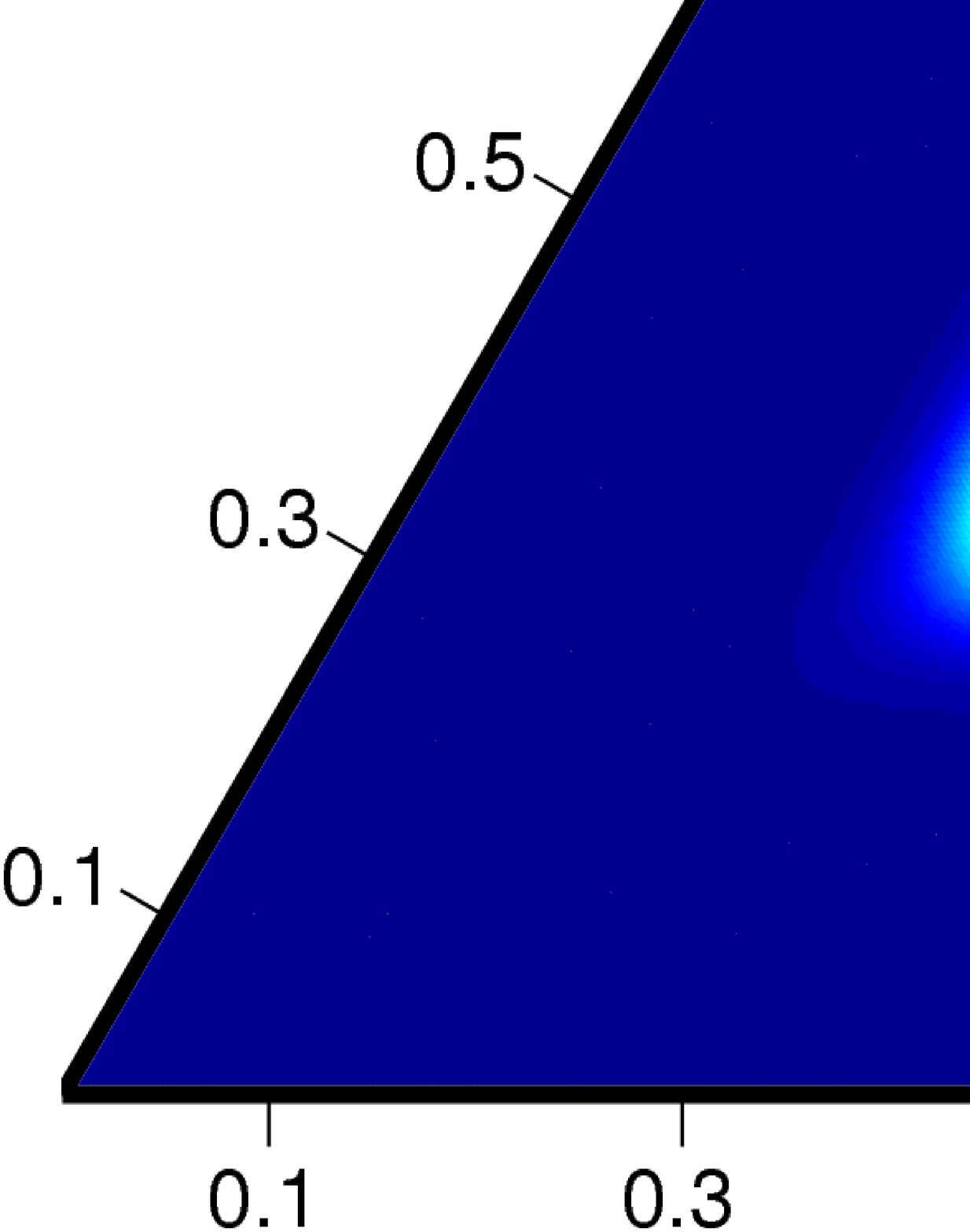} & 
\includegraphics[width=4.0 cm]{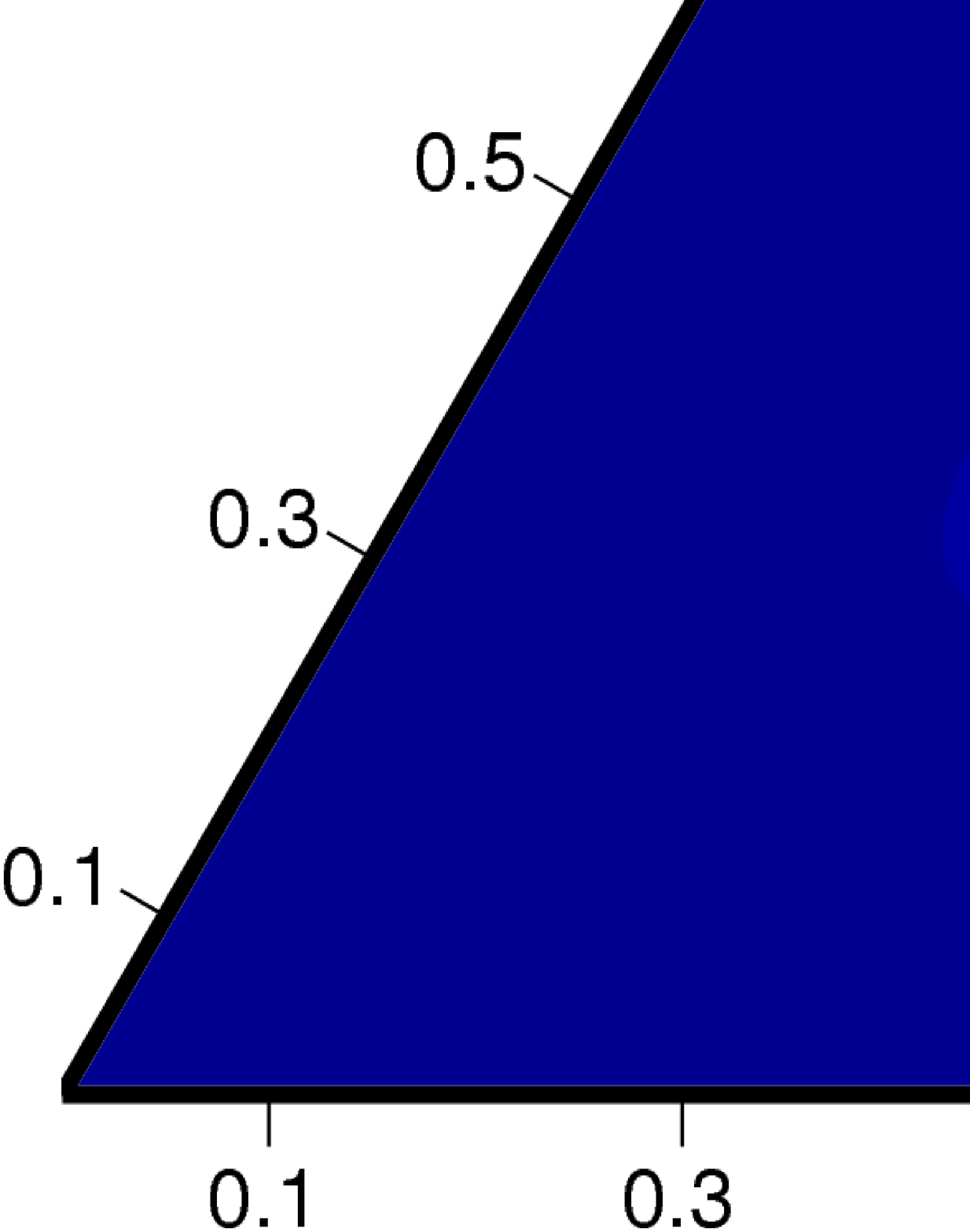}\\
\includegraphics[width=4.0 cm]{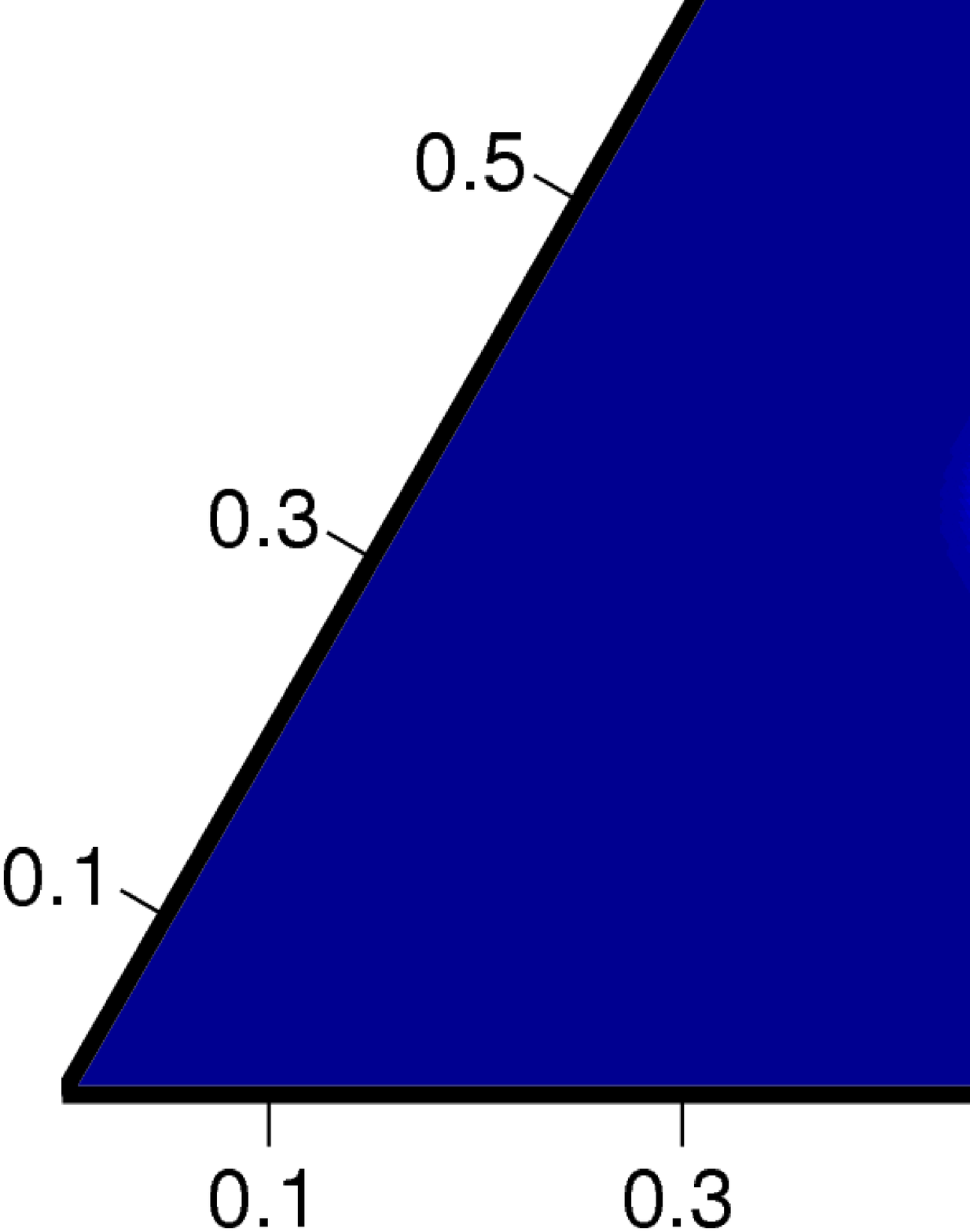} &
\includegraphics[width=4.0 cm]{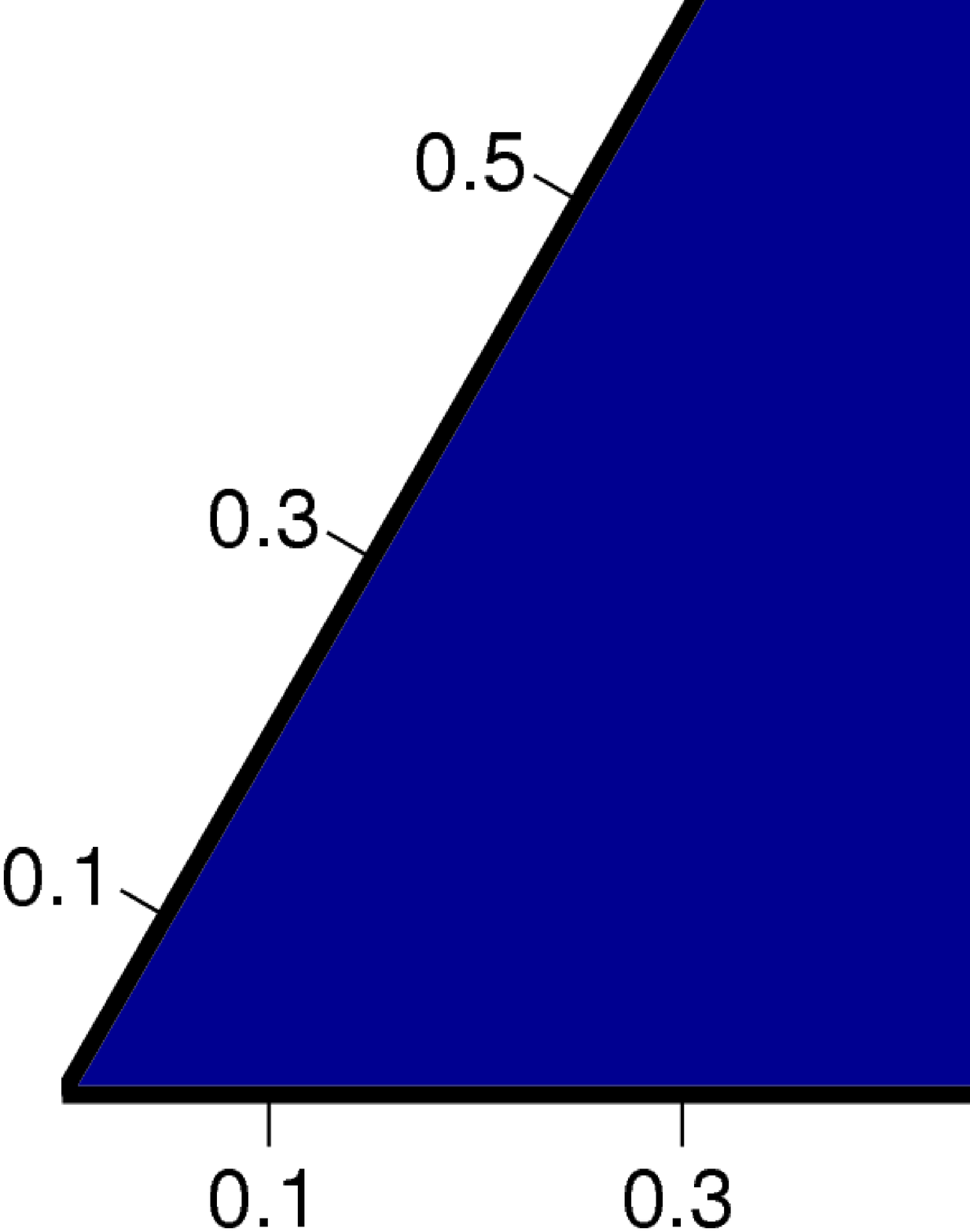} 
\end{tabular}
\caption{\label{trif2} Density plots of the probability distribution $P(\vec{\nu})$ for some values of the effective parameter $\tau$, Eq.~\eqref{tau}. Left: $N=300$. Right: $N=900$. The probability distribution is represented as described in the rightmost panel of Fig.~\ref{trif1}, but, in order to compare different total populations, the tick labels refer to $\nu_j/N$ instead of $\nu_j$. A white cross symbol signals the position of one of the three equivalent semiclassical ground-states (see text). The density plot is scaled with respect to  $P_{\rm M}= \max_{\vec{\nu}} P(\vec{\nu})$.}\end{centering}
\end{figure}

Figure \ref{trif2} shows the density plots of $P(\vec{\nu})$ for several values of the effective parameter $\tau$ defined in Eq.~\eqref{tau}. The representation is the same as in the rightmost panel of Fig.~\ref{trif1}, but in order to compare different total populations, the tick labels refer to $\nu_j/N$ instead of $\nu_j$. The left and right columns correspond to two different choices for the total population, $N=300$ and $N=900$, respectively. Note that the probability density undergoes a qualitative change as $\tau$ crosses the region of the semiclassical delocalization point, $\tau_{\rm d} = 0.25$. Below this threshold  $P(\vec{\nu})$ features three symmetrically positioned peaks. Notice that the probability peaks occur at the same positions as the three equivalent semiclassical ground states. The white cross symbol signals the position of one of such semiclassical ground-states, namely the one with $\nu_1=\nu_2$ and $\nu_3=N-2\,\nu_1$.
Owing to its symmetry, the three-modal distribution always produces a symmetric expectation value of the occupation numbers, $\langle \hat n_j\rangle = N/3$. However, a single measurement of $(\hat n_1, \hat n_2, \hat n_3)$ is extremely likely to produce a symmetrically broken  outcome very similar to the semiclassical result. 

As $\tau$ increases above $\tau_{\rm d}$, the three-modal probability distribution becomes monomodal. In this situation the symmetric expectation value $\langle \hat n_j\rangle = N/3$ is very similar to the outcome of a single measurement, in agreement with what happens at the semiclassical level.

Notice that there is a very narrow range  of  effective parameters in which the probability is essentially four-modal, featuring a central peak surrounded by three symmetrically positioned peaks. The value of $\tau$ such that the central peak has the same height as the peripheral peaks could be used to define the quantum counterpart of the delocalization threshold $\tau_{\rm d}$.
This situation is demonstrated in the density plots in the fourth row of Fig.~\ref{trif2}. These plots refer to slightly different values of $\tau$, depending on the different total populations. As the population increases, this quantum threshold approaches the semiclassical one, $\tau_{\rm d} = 0.25$. The remaining rows of  Fig.~\ref{trif2} refer to the same values of $\tau$.

The quantum threshold condition discussed above, $\max_{\vec{\nu}} P(\vec{\nu}) = P(\frac{N}{3},\frac{N}{3},\frac{N}{3})$, requires a tomography of the quantum state.
A substantially equivalent approach makes use of the average width defined in Eq.~\eqref{aqW}.

\subsection{Average width}
\begin{figure}
\begin{centering}
\includegraphics[width=8.2 cm]{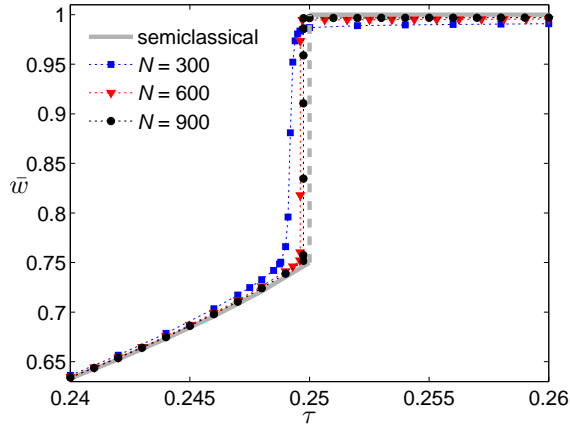}
\caption{\label{sqWL3} Average quantum width Eq.~\eqref{aqW} for a trimer containing 300, 600 and 900 bosons. The semiclassical result is also plotted for purposes of comparison.}
\end{centering}
\end{figure}

In this section we demonstrate how the measured average width for a quantum system, Eq.~\eqref{aqW}, reproduces the semiclassical results shown in Fig.~\ref{scW}. Once again we first of all consider the three-site lattice. Figure \ref{sqWL3} shows the average quantum width, Eq.~\eqref{aqW} for a lattice comprising $L=3$ sites and for different total boson populations, $N=300$, $600$ and $900$ particles. 
The semiclassical result appearing in Fig.~\ref{scW} is also plotted for purposes of comparison. Notice first of all that the range of $\tau$'s in Fig.~\ref{sqWL3} is much smaller than that in Fig.~\ref{scW}, and outside this range the quantum results corresponding to the three considered populations agree extremely well with the semiclassical findings. The differences between the three sets of quantum data and the semiclassical ones become perceptible in a narrow region surrounding the semiclassical localization/delocalization threshold, signalled by the vertical dashed gray line. Unlike its semiclassical counterpart, 
the average quantum width is a continuous function of the effective parameter $\tau$. However, its derivative becomes extremely large in a very narrow region in the proximity of the semiclassical threshold. The width of such region decreases with increasing boson population so that, in the limit $N\to\infty$ the semiclassical discontinuity corresponding to the catastrophe in the bifurcation pattern of the semiclassical ground state \cite{Buonsante_JPB_39_S77,Oelkers_PRB_75_115119} becomes apparent. The points on the almost vertical stretch of the quantum curves  correspond to  four-modal probability distributions, like in the fourth row of Fig.~\ref{trif1}. 
\begin{figure}
\begin{centering}
\includegraphics[width=8.2cm]{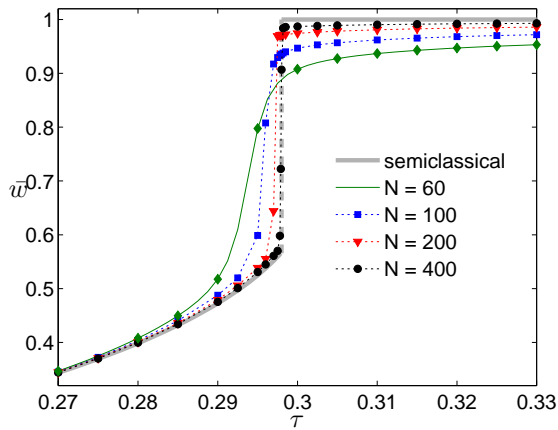}
\caption{\label{sqWL5} 
Average quantum width Eq.~\eqref{aqW} for a $L=5$ lattice  and several boson populations. The solid line black line for $N=60$ was obtained by means of Lanczos diagonalization, while all the  data points are {\it population} QMC results. The dotted lines are mere guides to the eye. The semiclassical result is also plotted for purposes of comparison.} 
\end{centering}
\end{figure}

The same qualitative behaviour is observed on lattices comprising four- and five-sites.  The data corresponding to several boson populations  are shown in Fig.~\ref{sqWL5} for the latter case. 
Even exploiting all of the system symmetries, the Hamiltonian matrix for the $L=5$ lattice becomes extremely challenging for the Lanczos algorithm around fillings of the order of 12 particles per site. We have nevertheless been able to tackle larger fillings by resorting to a stochastic approach, the so-called {\it population} QMC algorithm \cite{Iba_TJSAI_16_279}.
In Fig.~\ref{sqWL5} we also plot the results for $N = 60$ as provided by the Lanczos algorithm. The comparison with the data points for the same population demonstrates the reliability of our population QMC.

As already observed in the trimer case, Fig.~\ref{sqWL3}, outside a narrow range of $\tau$'s the quantum data corresponding to sufficiently large boson fillings overlaps very well with the semiclassical result. An incipient discontinuity is recognized in the proximity of the semiclassical threshold, which becomes more and more evident with increasing total population. The very same behaviour is oberved in the $L=4$ case, not shown.

The qualitative change at the transition taking place for lattice sizes larger than 5, evident in Fig.~\ref{scW}, is clearly recognizable also at the quantum level. 
\begin{figure}
\begin{centering}
\includegraphics[width=8.2 cm]{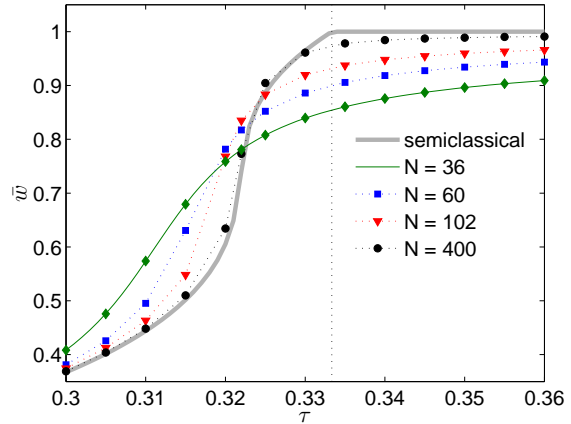}
\caption{\label{sqWL6} Average quantum width, Eq.~\eqref{aqW}, for a 6-site latice at different boson populations. The solid line corresponding to 30 (36?) bosons has been obtained by means of Lanczos exact diagonalization. The data points on top of it have been obtained by means of {\it population} QMC \cite{Iba_TJSAI_16_279}, as well as the data points for the larger populations. The dotted lines are mere guides to the eye. The semiclassical result is also plotted for purposes of comparison.  }
\end{centering}
\end{figure}
In order to illustrate this we consider the threshold situation, i.e. a lattice comprising $L=6$ sites. The data points in Fig.~\ref{sqWL6} --- also obtained  by means of the population QMC algorithm --- 
 illustrate the behaviour of the quantum width at several boson populations. The solid green line once again demonstrates the agreement between QMC and Lanczos algorithm at filling $N/L=6$. As the boson population is increased the data points for the quantum width get closer and closer to the semiclassical result which, at variance with smaller lattices, is continuous at the transition point.

Before concluding this paper we discuss in deeper detail the connection between the quantum theory of Hamiltonian \eqref{BH} and its semiclassical counterpart. 



\section{${\rm su}(L)$ coherent states}
\label{S:CS}
The DST fixed-point equations \eqref{DST} can be derived by assuming that  the
system is well described by a trial state of the form
\begin{equation}
\label{suL}
|\vec{\psi}\rangle = \frac{1}{\sqrt{N!}} \left(\sum_{j=1}^L \frac{\psi_j}{\sqrt{N}}\,a_j^\dag\right)^N |0\rangle\!\rangle.
\end{equation} 
This was first sketched in Ref.~\cite{Wright_PhisicaD_69_18}, where the trial state \eqref{suL} was referred to as {\it Hartree wave function}, and subsequently recast in terms of the {\it time-dependendent variational principle} in Ref.~\cite{Buonsante_PRA_72_043620}, where Eq.~\eqref{suL} was recognized as a su$(L)$ coherent-state (CS). 
After extremizing a suitable functional, both approaches result into Eq.~\eqref{DST}, except for a  correction factor $(N-1)/N$ cropping out in the interaction term. This comes about because  Eq.~\eqref{suL} is an eigenstate of the total number operator, $\hat N |\vec{\psi}\rangle = N |\vec{\psi}\rangle$, and is consistent with the expected absence of interaction when only one boson is present in the system. This correction provides a better agreement between quantum and semiclassical theories for small boson populations \cite{Buonsante_PRA_72_043620}, but can be safely neglected at most of the fillings considered in the present paper.   

The probability that a measurement of $\hat{\mathbf n}$ on the  CS~\eqref{suL} selects a Fock state  $|\vec{\nu}\rangle$, Eq.~\eqref{nu},
is given by
\begin{equation}
\label{nupsi}
P(\vec{\nu})=|\langle \vec{\nu}|\vec{\psi}\rangle|^2 = \frac{N!}{N^N} \prod_{j=1}^L \frac{\left(|\psi_j|^2\right)^{\nu_j}}{\nu_j!} 
\end{equation}
(see e.g. Ref.~\cite{Buonsante_JPA_41_175301}). Note that the above result applies when the total population of the Fock state is the same as the CS, 
 $\langle\vec{\nu}|\hat N|\vec{\nu}\rangle=\sum_j \nu_j=N$. By using standard methods (see e.g. Ref.~\cite{Huang_SM}) it is easy to appreciate that the above probability distribution  is sharply peaked at the Fock state best reproducing the set of semiclassical occupation numbers, $\nu_j \approx \langle\vec{\psi}|\hat n_j|\vec{\psi}\rangle = |\psi_j|^2 $ (see also the discussion about Fig.~\ref{trif3} below). 
Since the dynamical variables $\psi_j$ are determined by the DST equations \eqref{DST}, the CS breaks the translation symmetry below the localization threshold. That is, denoting $D$ the matrix producing a ciclic shift of the vector entries, $(D \vec{\nu})_j = \nu_{j+1} $, one gets $P(D^k\vec{\nu}) \neq P(\vec{\nu})$ for $k\neq L$.

In Ref.~\cite{Buonsante_PRA_72_043620} it was shown that, for sufficiently strong interactions, the above symmetry-breaking CS of  is well described by an uniform superposition of the lowest-energy states of the $L$ quasimomentum blocks of Hamiltonian \eqref{BH}.
Here we take a somewhat complementary standpoint, and  consider a uniform superposition of  $L$ equivalent, symmetry-breaking CS
\begin{equation}
\label{sCS}
|\vec{\phi}\rangle\!\rangle = \frac{1}{\sqrt{L}}\sum_{t=1}^L |D^t\vec{\phi}\rangle = \frac{1}{\sqrt{L}}\sum_{t=1}^L \hat D^{t}|\vec{\phi}\rangle 
\end{equation}
where the displacement quantum operator $\hat D$ is defined at the beginning of Sec. \ref{results}.
 The L (nonuniform) entries of the vector $\vec{\phi}$ are  normalized such that $\langle\!\langle\vec{\phi} |\vec{\phi}\rangle\!\rangle = \langle\vec{\phi} |\vec{\phi}\rangle = N^{-1}\sum_j |\phi_j|^2= 1$, are yet to be determined.
The optimal set of entries is such that the above symmetrized state attains the minimum energy. This entails
\begin{eqnarray}
\label{min1}
\frac{d}{d\phi_j^*}\left[ \langle\!\langle\vec{\phi} |\hat H|\vec{\phi}\rangle\!\rangle -\lambda  \left(\langle\!\langle\vec{\phi} |\vec{\phi}\rangle\!\rangle-1\right)\right] &=& 0\\
\label{min2}
\frac{d}{d\lambda}\left[ \langle\!\langle\vec{\phi} |\hat H|\vec{\phi}\rangle\!\rangle -\lambda  \left(\langle\!\langle\vec{\phi} |\vec{\phi}\rangle\!\rangle-1\right)\right] &=& 0
\end{eqnarray} 
where $\lambda$ is a Lagrange multiplier enforcing the constraint on the norm.
More explicitly this means
\begin{widetext}
\begin{eqnarray}
\lambda  \sum_r \Pi^{N-1}_{r} \phi_{h +r}
&=& 
- \sum_r 
\left[\,  \frac{N-1}{N}  \Pi^{N-2}_r \,U \phi^*_{h} \phi_{h+r}^2  
+J \, \Pi^{N-1}_r \, \left ( \phi_{h+r-1} \, +  \phi_{h+r+1}  \right ) 
\right] 
\nonumber\\
\label{minim1}
&-& \frac{N-1}{N}  \sum_r   \phi_{h+r}\sum_s 
\left[ 
\frac{N-2}{N}  \Pi^{N-3}_r 
\,\frac{U}{2} \, (\phi^*_{s})^2 \phi_{s+r}^2 
+J   \, \Pi^{N-2}_r \, \left( 
\phi_{s+r-1}  +  \phi_{s+r+1} \right)\, \phi^*_{s}\,
\right],
\end{eqnarray}
\end{widetext}
and
\begin{equation}
\label{minim2}
 \sum_r \Pi_r^N = 1
\end{equation}
where we introduced the shorthand notation $\Pi_r~=~N^{-1}\sum_s \phi_s^* \phi_{s+r} $.
Since  nonuniform $\vec{\phi}$'s yield $\Pi_0 > \Pi_k $, the only possible solution of Eq.~\eqref{minim2} in the $N\to \infty$ limit is $\Pi_0=1$, i.e. the usual normalization condition $\sum_j |\phi_j|^2=N$. This entails that $\Pi_r^N \stackrel{N\to\infty}{\longrightarrow} \delta_{r, 0}$, which
considerably simplifies Eq.~\eqref{minim1}
\begin{equation}
\label{minim1b}
(\lambda-{\cal E}) \phi_h = -\frac{N-1}{N} U |\phi_h|^2 \phi_h - J(\phi_{h+1}+\phi_{h-1}) 
\end{equation}
where ${\cal E} = \langle\vec{\phi}|\hat{H}|\vec{\phi}\rangle$. Comparing Eqs.~\eqref{DSTfp} and Eq.~\eqref{minim1b} shows that, in the large population limit, the parameters $\vec{\phi}$ to be plugged into the symmetrized CS \eqref{sCS} do not need to be obtained from the (numerically demanding) minimization inherent in Eqs.~\eqref{min1} and \eqref{min2}, but can be replaced by the (much more easily determined) normal modes of the dynamical DST equations \eqref{DST}, i.e. the solutions of Eq.~\eqref{DSTfp}. The comparison of Eqs.~\eqref{DSTfp}, \eqref{tau} and \eqref{minim1b} shows that semiclassical parameter  corresponding to Eq.~\eqref{minim1b} must be rescaled by a factor $\frac{N}{N-1}$.

\begin{figure}
\begin{centering}
\begin{tabular}{ccc}
\includegraphics[width=2.8 cm]{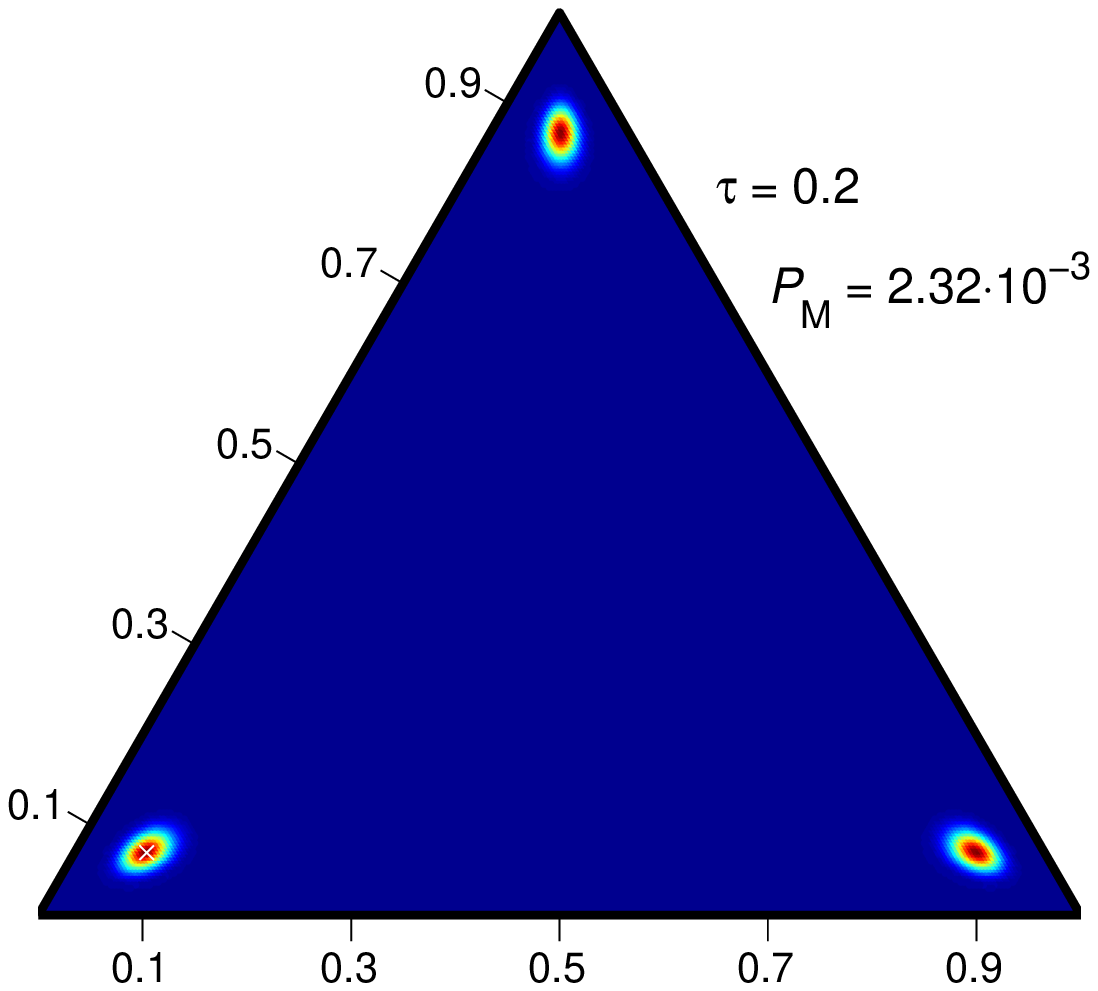} & 
\includegraphics[width=2.8 cm]{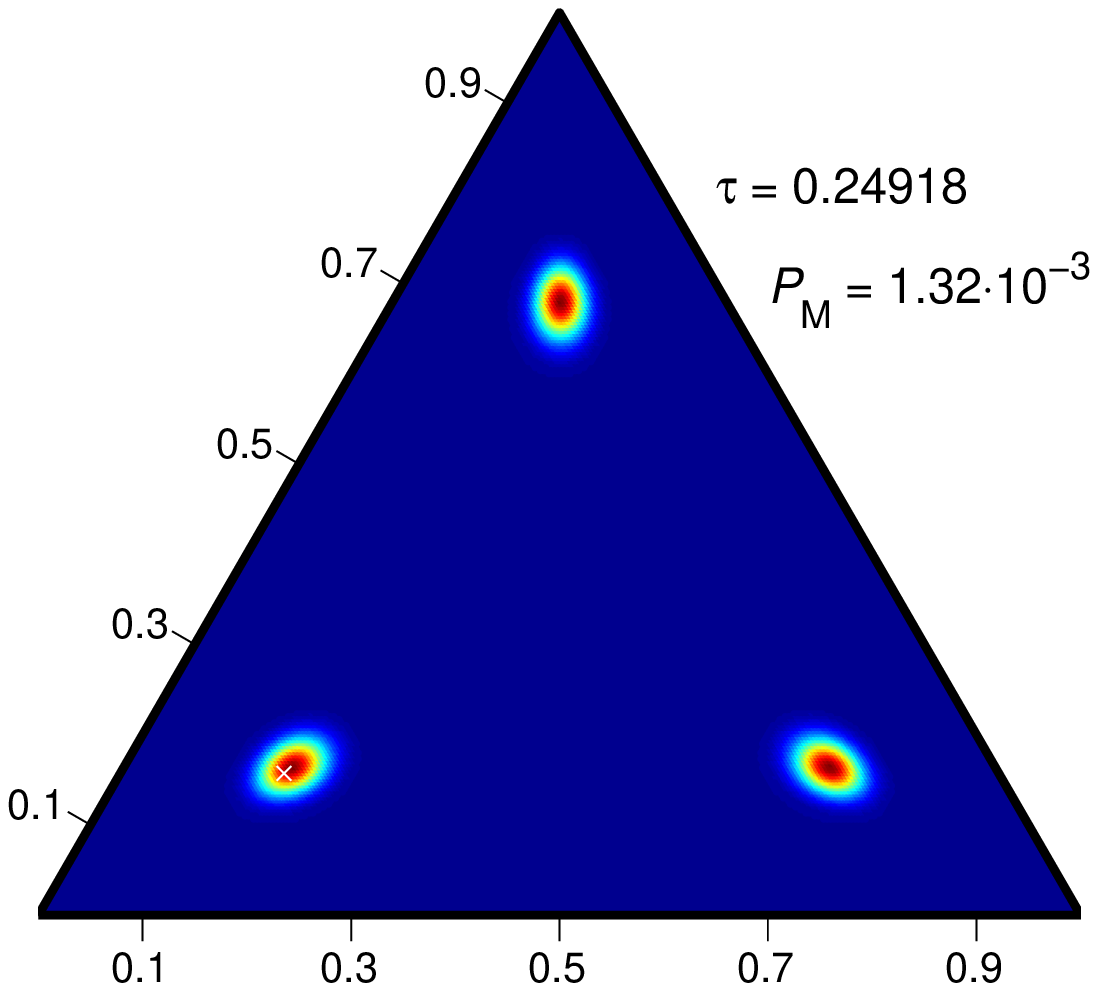} & 
\includegraphics[width=2.8 cm]{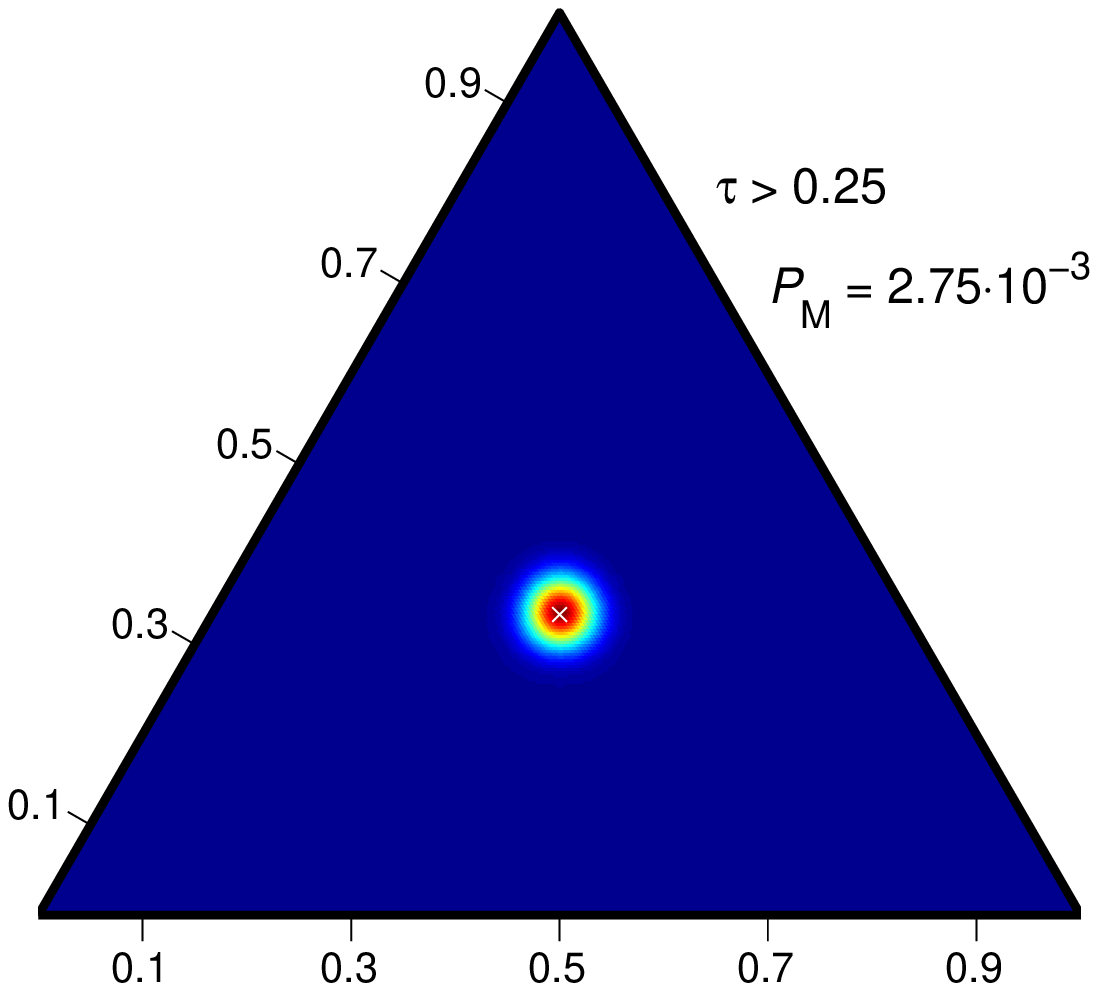}  
\end{tabular}
\caption{\label{trif3} Density plots of the probability distribution $P(\vec{\nu})=|\langle\vec{\nu}|\vec{\phi}\rangle\!\rangle|^2$ corresponding to the symmetrized coherent state in Eq.~\eqref{sCS} for a trimer containing $N=300$ bosons.
We choose the same values of the effective parameter $\tau$, Eq.~\eqref{tau} as in the first, third and fourth panel from top  in the left column of Fig.~\ref{trif2}.  A white cross symbol signals the position of one of the three equivalent semiclassical ground-states (see text). The density plot is scaled with respect to  $P_{\rm M}= \max_{\vec{\nu}} P(\vec{\nu})$.}\end{centering}
\end{figure}

In Fig.~\ref{trif3} we show some density plots for the probability distribution  $P(\vec{\nu})=|\langle\vec{\nu}|\vec{\phi}\rangle\!\rangle|^2$ corresponding to the symmetrized coherent state in Eq.~\eqref{sCS} for a three-site lattice containing $N=300$ bosons. The  parameters $\vec{\phi}$ are not obtained by solving Eqs.~\eqref{min1} and \eqref{min2}. Instead, in keeping with the above discussion, we use the solution of  the DST fixed-point equation \eqref{DSTfp}, given the fairly large boson filling. The parameter $\tau$ controlling the relative strength of the tunneling and  interaction energy, Eq.~\eqref{tau}, has been chosen as in the first, third and fourth panels in the left column of Fig.~\ref{trif2}, respectively. As we mentioned earlier, the probability density \eqref{nupsi} for the su$(N)$ coherent state \eqref{suL} is strongly peaked at the occupation numbers $\{\nu_j\}$ best reproducing the semiclassical local populations $\{|\phi_j|^2\}$. It is then clear that, in the localized regime, the density of the symmetrized state \eqref{sCS} is qualitatively similar to that of the quantum state analyzed in Fig.~\ref{trif2}. It features three peaks, each relevant to one of the Fock states corresponding to the tree equivalent, symmetry-breaking solutions of Eq.~\eqref{DSTfp}. For  sufficiently large interactions (small $\tau$'s) the similarity is striking, as it was already pointed from a different perspective in Ref.~\cite{Buonsante_PRA_72_043620}.
 The differences between the quantum ground state in  Fig.~\ref{trif2} and the symmetrized CS in  Fig.~\ref{trif3} become evident in the vicinity of the semiclassical threshold. In particular, at variance with the former, the latter is structurally unable to give rise to a four-modal distribution, as it is clear after comparing the central panel of  Fig.~\ref{trif3} and the third panel from top in the left column of Fig.~\ref{trif2}. 

Above the delocalization threshold there is no need of symmetrizing the CS in Eq.~\eqref{suL}, because the ground-state solution of Eq.~\eqref{DSTfp} becomes translation invariant, $\phi_j = \sqrt{N/M}$, which makes  the  CS and the corresponding probability density  $\tau$-independent. As it is clear from the comparison between the rightmost panel in Figs.~\ref{trif3} and the bottom panel in the left column of Fig.~\ref{trif2}, the similarity between the CS and the quantum ground state is already rather satisfactory at $\tau=1$. It improves with increasing $\tau$, and becomes perfect in the noninteracting limit.

\begin{figure}
\begin{centering}
\includegraphics[width=8.2 cm]{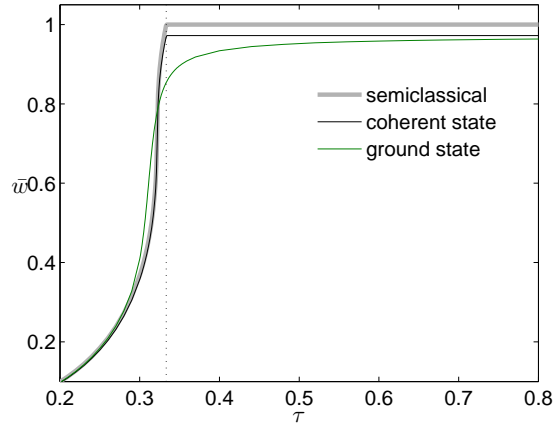}
\caption{\label{sqWL6cs} Comparison of the (squared) width of the  semiclassical ground state, the coherent state and the exact quantum ground-state for a six-site lattice. The latter results refer to total population of $N=36$ bosons. a  Note that the three curves overlap in the small-$\tau$ region, and that the coherent state provides a better approximation in the large-$\tau$ limit at the relatively small filling considered. } 
\end{centering}
\end{figure}

We conclude by observing that the (quantum) width of  CS \eqref{sCS} as calculated with Eq.~\eqref{aqW} coincides with the semiclassical value Eq.~\eqref{w2} only in the large-$N$ limit. Straightforward calculations show indeed that 
\begin{equation}
\bar w({|\phi\rangle\!\rangle})= \frac{N-1}{N} w\left(\left\{|\phi_j|^2\right\}\right)
\end{equation}
The population-dependent prefactor in the previous equation correctly allows of the expected delocalization when one single boson is present in the system. Also, it provides a better agreement with the exact result in the large-$\tau$ limit. As we discuss above, the same prefactor affects the semiclassical parameter $\tau$ to be used in Eq.~\eqref{minim1b}, defined in Eq.~\eqref{tau}. This correction makes it so that the width of the symmetrized CS is closer to that of the quantum ground state in the small-$\tau$ (large interaction) regime. 
This should be clear from Fig.~\ref{sqWL6cs}, and shows that the su$(L)$ CS approach to the semiclassical theory  is more effective in capturing the effects arising from finite size than the simple replacement of lattice operators by C-numbers used in deriving Eq.~\eqref{DST}. 

\section{Conclusions}
In this paper we perform a systematic analysis of ground-state properties  of a system of attractive lattice bosons, highlighting the correspondences between the (inherently linear) quantum theory for this system and its nonlinear semiclassical counterpart.
Our analysis relies on the introduction of a suitable measure of the width of 
the symmetry-breaking {\it soliton-like} ground state characterizing the nonlinear semiclassical theory in the large interaction limit, which is then readily transported to the quantum level.
This quantity allows us to perform a systematic comparison between the semiclassical and quantum ground state, exposing  striking similarities and significantly extending the discussion in Ref. \cite{Javanainen_PRL_101_170405}. 
On the one hand, the comparison of the semiclassical localized state and its quantum counterpart is made more quantitative. On the other hand, we extend the parameter range to include the localization/delocalization transition occurring in the semiclassical nonlinear theory, and show that it has a clear correspondent at the quantum level. In particular, we demonstrate that the change in the semiclassical bifurcation pattern is maintained also quantum-mechanically.
Our analysis is enriched by a detailed investigation of the three-site case, which makes it possible to visualize directly the structure of the quantum ground state. We also include a somewhat rigorous discussion of the relation between the quantum theory for attractive lattice bosons and its semiclassical version, related to the discrete self-trapping equations. This highlights a {\it finite-population} effect.

\end{document}